\documentstyle[12pt]{article}

\def\av#1{\left<#1\right>}
\def\qav#1{\left[#1\right]_{\rm av}}
\def\sign{\mathop{\rm sign}}
\title{FITNESS LANDSCAPES AND EVOLUTION\thanks{
Lectures given at the NATO ASI on Physics of Biomaterials:
Fluctuations, Self-Assembly, and Evolution, Geilo (Norway),
March 27-April 6, 1995.}}
\author{LUCA PELITI\\
Dipartimento di Scienze Fisiche and Unit\`a INFM\\
Universit\`a ``Federico II'', Mostra d'Oltremare, Pad.~19\\
I-80125 Napoli (Italy)\thanks{Associato INFN, Sezione di
Napoli. E-mail: {\tt peliti@na.infn.it}}
}

\begin{document}
\maketitle
\begin{abstract}
The concept of fitness is introduced, and a simple derivation
of the Fundamental Theorem of Natural Selection (which states
that the average fitness of a population increases if its
variance is nonzero) is given. After a short discussion of the
adaptative walk model, a short review is given of the
quasispecies approach to molecular evolution and to the
error threshold. The relevance of flat fitness landscapes
to molecular evolution is stressed. Finally a few examples
which involve wider concepts of fitness, and in particular
two-level selection, are shortly reviewed.
\end{abstract}
\section{Fitness and the fundamental theorem of natural selection}
The term ``fitness''\index{fitness} derives from the phrase
``survival of the fittest'' that the philosopher
Herbert Spencer\index{Spencer, H.}
suggested to use instead of ``natural selection''.
In the struggle made by the evolutionary theorists to avoid
the tautology lurking in the phrase, the term has
been twisted to several meanings. R.~Dawkins~\cite{Dawkins82}
\index{Dawkins, R.}
distinguishes no less than {\em five\/} different
meanings to the word in the evolutionary literature.
{}From the point of view of model building,
the most convenient meaning---and the one we shall
adopt---is however the following:
\begin{quote}
The fitness of an {\em individual\/} is {\em proportional\/}
to the {\em average\/} number of offspring it may have in
the given environment.
\end{quote}
In this definition, fitness is assigned to individuals
rather that to genes or to groups of individuals. It is further
assumed that reproduction takes place via a stochastic process,
and that, in a given population, the average numbers
of (immediate) offspring of two individuals have
the same ratio as their fitnesses: therefore, only
ratios of fitnesses have a well-defined meaning, and not their
absolute value.

Let us consider a population formed by a certain number
of individuals, whose inheritable characteristics
(genotype)\index{genotype} are summarized by the variable $\sigma.$
Let us further assume that the population reproduces asexually, that
the offspring of an individual have the same genotype
as the parent, and finally that the number of
offspring is exactly proportional to the fitness
of the parent: briefly, let us neglect {\it mutations\/}\index{mutation}
in the genotype and {\it fluctuations\/} in the number of
offspring.

We can thus write down an equation expressing the
number $n_t(\sigma)$ of individuals carrying the genotype
$\sigma$ at generation $t+1,$ given the same quantity
at generation $t,$ assuming that the fitness $A(\sigma)$
of an individual is a function only of its genotype:
\begin{equation}
n_{t+1}(\sigma)=\frac{1}{{\cal Z}_t} A(\sigma)n_t(\sigma),\label{fund:eq}
\end{equation}
where ${\cal Z}_t$ is a proportionality constant.
In order to simplify the argument we have also assumed
that the generations are {\em nonoverlapping}, i.e.,
that al individuals, once reproduced, die.

The total number ${\cal N}_t$ of individuals in the population
at generation $t$ is given by
\begin{equation}
{\cal N}_t=\sum_{\sigma}n_t(\sigma).
\end{equation}
We define the {\em population average\/} $\av{Q}_t$ at
generation $t$ of a quantity
$Q(\sigma),$ which depends only on the genotype $\sigma,$ in the
following way:
\begin{equation}
\av{Q}_t=\frac{1}{{\cal N}_t}\sum_{\sigma}Q(\sigma)n_t(\sigma).
\end{equation}

We can now prove that the average fitness $\av{A}$ always
{\em increases}, unless all individuals have the same fitness.
We have in fact:
\begin{equation}
\av{A}_{t+1}=\frac{1}{{\cal N}_{t+1}}\sum_{\sigma}A(\sigma)n_{t+1}(\sigma)
=\frac{1}{{\cal N}_{t+1}{\cal Z}_t}\sum_{\sigma}A^2(\sigma)n_{t}(\sigma).
\label{Fisher:eq}
\end{equation}
On the other hand, one has
\begin{equation}
{\cal N}_{t+1}=\frac{1}{{\cal Z}_t}\sum_{\sigma}
A(\sigma)n_{t}(\sigma)=\frac{{\cal N}_t}{{\cal Z}_t}\av{A}_{t}.
\end{equation}
Therefore
\begin{equation}
\av{A}_{t+1}\av{A}_{t}=\av{A^2}_{t}\ge \left(\av{A}_{t}\right)^2,
\end{equation}
and the equality holds only if all individuals in the population
have the same fitness. In fact, the larger the variance in
the fitness, the faster its average grows.

This result is a simplified version of the Fundamental Theorem
\index{Fundamental Theorem of Natural Selection} of
Natural Selection due to R.~Fisher~\cite[p.~22ff]{Fisher58}.
\index{Fisher, R.} Some authors have
considered it as the key point of difference between the
living and the inorganic world.
As K.~Sigmund puts it~\cite[p.~108]{Sigmund93}:\index{Sigmund, K.}
\begin{quote}
So we see, in physics, disorder growing inexorably
in systems isolated from their surroundings; and in biology,
fitness increasing steadily in populations
struggling for life. Ascent here and degradation
there---almost too good to be true.
\end{quote}

In fact, the result depends on many unrealistic assumptions.
Let alone the complications introduced by sex,
\index{sex} which lead to
maddeningly complex behavior, let us focus on the effects
of mutation: on that set of causes which makes
offspring different from their parent, even among
bacteria.

We all know that genetic information is carried by the\index{DNA}
DNA, in the form of a sequence of nucleotide\index{nucleotide}
bases, which
belong to four different types: {\tt A} adenine and
{\tt G} guanine (purines); {\tt T} thymine and {\tt C}
cytosine (pyrimidines). In the double helix of DNA they
are found in matching pairs: {\tt A-T} and {\tt G-C}.
During the replication, it may happen that the replication
mechanism, which associates one of the ``old'' strands
to the ``new'' ones, stumbles in some errors. These errors can
be divided in a few classes:\index{mutation}
\begin{description}
\item[Point mutations:] Substitution of one nucleotide base
to another. They can be divided into two classes:\hfill\break
{\em Transitions\/} (the most common): substitution of
one purine by the other, or of one pyrimidine by the other;\hfill\break
{\em Transversions}, in which a purine is replaced
by a pyrimidine and viceversa.
\item[Insertions and deletions:] They correspond to the introduction
of new bases in the strand or in their omission respectively.
In the case of sequences coding for a protein,\index{protein}
these mutations
are often fatal, since they entail a frame shift in the translation
into proteins, unless they occur by threes.
\item[Major rearrangements:] In this class one considers the
insertions (or deletion) of comparatively long sequences. This
is the case, e.g., of the
\index{transposable elements}{\em transposable elements\/} which are
known to move easily from one place to another in the genotype.
A subclass of special interest is {\em gene doubling.}
\end{description}

These processes do not have the same probability. If one considers
two genotypes, i.e., two different nucleotide sequences in DNA,
one may introduce a notion of distance between them by
considering the probability of the most likely
mutation path connecting them.
This notion of distance (metrics) has a rather immediate
evolutionary meaning, but is most often quite difficult to compute.
For the sake of definiteness I shall consider in the following
only {\em point mutations}, and I shall assign the same probability
to transitions and to transversions.
In this case all DNA\index{DNA} sequences
which may be connected to each other have the same length, and
their distance is equal to the number of points in the sequence
in which different bases are found: this is known as the {\em
Hamming distance.}\index{Hamming distance}

To summarize: we have defined a {\em genotype space\/}
\index{genotype space} as the space
of all sequences $\sigma$ of a given length which can be built with
the four-letter alphabet {\tt ATGC}. This space is endowed with
a metrics, defined by the Hamming distance, i.e., by the
number of corresponding positions in the sequence in which
different nucleotide pairs are encountered. If we now assign
to each such sequence its fitness value (assuming that the fitness
of an individual is a function only of its genotype), we obtain
a {\em fitness landscape.} The phrase goes back to \index{Wright, S.}
S.~Wright~\cite{Wright32}, but the concept can already be found in
Fisher's work in the 'twenties.\index{fitness landscape}

The Fundamental Theorem therefore implies that populations
move on fitness landscapes striving to climb {\em up\/} their
peaks, while we are accostumed to physical systems
rolling {\em down\/} the slopes towards the points of smallest energy.
In this sense, fitness plays in evolutionary theory a role similar to
energy in mechanics.

\section{Adaptative walks}
The Fundamental Theorem intimates in fact that the population rapidly
reaches the maximum fitness\index{fitness}
of all individuals that are already
present in it. Higher values of the fitness can only arise if
there are mutations. If mutations are rare, one can think of
a regime in which mutants arise from time to time and, if they
correspond to higher fitness, ``draw'' the population to the
new fitness value. This justifies the evolutionary model known
as the {\em Adaptative Walk\/} \cite{Kauffman93}.\index{adaptative
walk}\index{Kauffman, S. A.}

In order to simplify the discussion, we shall consider
from now on a genotype written in a two-letter alphabet.
The conclusions that we shall draw can be easily translated,
in principle, in the four-letter alphabet of real life.
We shall denote the two letters by $\{-1,+1\},$ and\index{genotype}
describe the genotype $\sigma$ by a collection of $N$
binary variables (units): $\sigma=(\sigma_1,\sigma_2,
\ldots,\sigma_N),$ where $\sigma_i=\pm 1,$ $\forall i.$
The space of these genotypes
is the hypercube in $N$ dimensions, whose $2^N$
vertices correspond to the genotypes, and the
Hamming\index{Hamming distance}
distance between genotypes $\sigma$ and $\sigma'$ is given by
\begin{equation}
d_{\rm H}(\sigma,\sigma')=\frac{1}{2}\sum_{\sigma}\left(1
-\sigma_i\sigma'_i\right).
\end{equation}
We shall also consider an equivalent measure of the
similarity or dissimilarity of genotypes, namely
the {\em overlap\/}\index{overlap}
$q,$ central to the theory\index{spin glass}
of spin glasses~\cite{Mezard87}, and defined by
\begin{equation}
q=\frac{1}{N}\sum_{i=1}^N\sigma_i\sigma'_i=1-\frac{2 d_{\rm H}
(\sigma,\sigma')}{N}.
\end{equation}
The overlap between identical genotypes is equal to $1,$
and it decreases as the Hamming distance increases.
Two completely independent genotypes will be different,
on average, in half of their units, and the corresponding
overlap will be close to zero.

We can now define the Adaptative Walk model\index{adaptative walk}
\cite{Kauffman87}, \cite[p.~39--40]{Kauffman93}.
To each genotype $\sigma$ is associated
its fitness $A(\sigma).$ One assumes that the population
is characterized by a single genotype $\sigma(t)$ at each
generation $t.$ The initial genotype $\sigma(0)$ is chosen
at random. Given the genotype $\sigma(t),$ the next genotype
is chosen according to the following procedure:
\begin{itemize}
\item[(i)] One changes sign to one of the units of the genotype
$\sigma(t),$ chosen at random; in other words, one chooses
at random one of the $N$ vertices of the hypercube
closest to $\sigma(t);$ one thus obtains a tentative
genotype $\sigma'(t);$
\item[(ii)] If $A(\sigma'(t))>A(\sigma(t)),$ then
$\sigma(t+1)=\sigma'(t);$ otherwise $\sigma(t+1)=\sigma(t).$
\end{itemize}

This procedure is reminiscent of a zero-temperature\index{Monte Carlo}
Monte-Carlo dynamics, where the Hamiltonian is a decreasing
function of the fitness $A(\sigma).$ Evolution is bound
to finish on a local fitness\index{fitness} maximum. More explicit
predictions can only be made when more properties of the
fitness landscape are known.\index{fitness landscape}

We do not know in general the intricate conditions
which determine the fitness of a given species as a function
of its genotype: we can only expect the fitness landscape
to be rather irregular and complicated. It has been
suggested \cite{Anderson83}\index{Anderson, P. W.}
to represent a given fitness landscape as a
realization of a random function.

A rather general class of random functions defined
on the $N$-dimensional hypercube has been introduced by
B.~Derrida in the context of spin-glass theory \cite{Derrida81}. It
is defined by the expression\index{spin glass}\index{Derrida, B.}
\begin{equation}
A^{(p)}(\sigma)=\sum_{\{i_1,i_2,\ldots,i_p\}}
J_{\{i_1,i_2,\ldots,i_p\}}\sigma_{i_1}\sigma_{i_2}\ldots\sigma_{i_p},
\end{equation}
where the sum runs over all different subsets
of $n$ indices, and for each such subset, the
coefficient $J_{\{i_1,i_2,\ldots,i_p\}}$ is an independent,
identically distributed, real random variable.
This model is known as the $p$-spin model
in spin-glass theory.
A slightly different set of random functions has
been independently introduced by S.~A.~Kauffman
in the context of Adaptative Walks \cite{Kauffman87},
\cite[p.~54--62]{Kauffman93},
where it is known as the $NK$-model.\index{Kauffman, S. A.}
\index{$NK$-model}

The simplest case is of course $p=1.$ In this case
the maximum fitness is reached for the single
genotype\index{genotype} $\sigma^*,$ satisfying
\begin{equation}
\sigma^*_i=\sign J_i.
\end{equation}
Moreover, since there are no local maxima but $\sigma^*,$
it is possible to reach this maximum simply by flipping
one unit after the other in the good direction.
Evolution is a simple matter in this ``Fujiyama landscape'',
\index{Fujiyama}
as it has been called, because it is never necessary
to undo the progress already made in order to go forward.

However, as soon as we go to $p>1,$ thing become much
more complicated.

Two properties of the landscape are strictly related: the
frequency of local fitness maxima (peaks), and the correlation
(or its contrary, ``ruggedness'') of the landscape.
We say that a landscape is {\em rugged\/}
if the value $A^{(n)}(\sigma)$ of the fitness changes a great deal
when the genotype $\sigma$ changes sligtly.
A measure of the ruggedness of the landscape is provided
by the correlation function
$C(\sigma,\sigma')=\qav{A^{(p)}(\sigma)A^{(p)}(\sigma')},$ where
$\sigma$ and $\sigma'$ are two different genotypes with
overlap $q,$ and the average $\qav{}$ is taken over the probability
distribution of the coefficients $J.$\index{correlation}
In the ``thermodynamic limit'' $N\to\infty,$ this
quantity is equal, with probability one,
to the average of the product $A^{(p)}(\sigma)A^{(p)}(\sigma')$
taken over all genotype pairs with overlap\index{overlap} equal to $q.$
If this correlation function decays slowly with
decreasing overlap $q,$ the landscape is smooth;
otherwise, it is rugged.\index{rugged fitness landscape}
The more rugged the landscape, the larger the frequency
of local optima.

Let us assume, with Derrida~\cite{Derrida81},
that the distribution of the coefficients $J$
is a Gaussian of mean zero, and variance
equal to $J_0^2\,p!/(2N^{p-1}).$ One can then prove
that the probability density $P_{\sigma}(E)$ that the fitness
$A^{(p)}(\sigma)$ of any given genotype $\sigma$
is equal to $E$ is also a Gaussian:
\begin{equation}
P_{\sigma}(E)=\qav{\delta\left(
A^{(p)}(\sigma)-E\right)}\propto\exp\left[-\frac{E^2}{NJ_0^2}\right].
\end{equation}
The properties of the landscape\index{fitness landscape} can
be read off the joint probability distribution
function
\begin{eqnarray}
P_{\sigma\sigma'}(E,E')&=&\qav{\delta\left(A^{(p)}(\sigma)-E\right)
\delta\left(A^{(p)}(\sigma')-E'\right)}\nonumber\\
&\propto&\exp\left[-\frac{(E+E')^2}{2NJ_0^2(1+q^p)}
-\frac{(E-E')^2}{2NJ_0^2(1-q^p)}\right],
\end{eqnarray}
where $q=(1/N)\sum_i\sigma_i\sigma'_i$ is the overlap
of the two configurations.
The correlation function $C(\sigma,\sigma')$ then reads
\begin{equation}
C(\sigma,\sigma')=\qav{A^{(p)}(\sigma)A^{(p)}(\sigma')}=
\frac{1}{2}NJ_0^2q^p.
\end{equation}
It decays more and more rapidly as $p$ increases.
As $p$ increases, therefore,
the landscape becomes more and more ``rugged''.
At the same time, the number of extrema becomes
larger and larger. Already for $p=2,$ which corresponds
to the Sherrington-Kirkpatrick model of spin glasses,
\index{Sherrington, D.}\index{Kirkpatrick, S.}\index{spin glass}
this number increases exponentially with $N.$
Therefore, it becomes more and more likely that
the adaptative walk, starting from an arbitrary
initial genotype, ends up in a local fitness\index{genotype}
\index{fitness}
maximum instead of the absolute one.

Eventually, as $p\to\infty$
one obtains, whenever $\left|q\right|<1,$
\begin{equation}
P_{\sigma\sigma'}(E,E')\sim P_{\sigma}(E)P_{\sigma'}(E').
\end{equation}
This is known as the ``rugged landscape'' limit, in which the
\index{rugged fitness landscape}
fitnesses corresponding to different genotypes
are independent random quantities.
Adaptative walks in this limit have been thoroughly discussed
by Kauffman and Levin \cite{Kauffman87}\index{Kauffman, S. A.}
\index{Levin, S.}
and more recently analyzed by
H.~Flyvbjerg and B.~Lautrup \cite{Flyvbjerg92}.
\index{Flyvbjerg, H.}\index{Lautrup, B.}

Let us therefore consider adaptative walks\index{adaptative walk}
in a landscape in which the fitness $a=A(\sigma)$ of each
different genotype $\sigma$ is an independent\index{genotype}
random variable, with a given probability distribution
function $p(a).$
Several important properties of this model can
be obtained almost immediately \cite[p.~47--52]{Kauffman93}:
\begin{itemize}
\item {\em The probability that a given genotype $\sigma$ is
a local fitness optimum is equal to\/} $1/(N+1).$ This can
be simply evaluated in terms of the cumulative distribution
function $\Phi(a)=\int_{-\infty}^ada'p(a'),$
namely, the probability that the
fitness of a given genotype is smaller than $a.$
Calling $P_N$ the probability that a given genotype
is a local maximum, we have indeed:
\begin{equation}
P_N=\int_{-\infty}^{\infty}da\,p(a)\;\Phi(a)^N=\frac{1}{N+1}.
\end{equation}
\item {\em A walk leading to a local optimum will touch on average
$\simeq\log_2 N$ different genotypes.} In fact, since there
are no correlations between the
value of the fitness\index{fitness} of one
genotype and that of its next (fitter) mutant---except, of course,
that it is larger---the value of $1-\Phi$ will be halved, on average,
at each mutation step. The previous result tells us that the walk
will stop when $1-\Phi\simeq 1/N.$ Calling $\ell$ the length
of the walk, we thus have $1-2^{-\ell}\simeq 1-1/N,$ hence
$\ell\simeq\log N/\log 2.$
\item {\em The expected time $T$ needed to reach an optimum
is proportional to $N.$} The idea is that the waiting time at each
step is inversely proportional to the probability that
any given mutant is fitter, i.e., to $1-\Phi(a).$
Thus the waiting time {\em doubles\/} at each step.
\index{waiting time}
We obtain therefore, roughly
\begin{equation}
T\simeq\sum_{k=0}^{\ell-1}2^k= 2^{\ell}-1=N-1.
\end{equation}
\item {\em The expected fitness $a^*$ of local optima satisfies
the equation $\Phi(a^*)=1-1/N.$} This result has an important
consequence, named by Kauffman ``the complexity catastrophe''.
As $N$ increases, it is reasonable to assume that the ``typical''
values of $a$ increase like some power of $N.$ On the other
\index{complexity catastrophe}
hand, $1-\Phi(a)$ usually decreases faster than any power
as $a\to\infty.$ Therefore, as $N$ increases, the fitness
values of the local optima become closer and closer to the
``typical'' values.
\end{itemize}

Before going on, let us emphasize, following \cite{Kauffman93},
that similar results are expected to hold qualitatively
also for adaptative walks\index{correlated fitness landscapes}
on correlated (smoother) landscapes.
Let us consider such a landscape, and assume that the correlation
function $C(\sigma,\sigma')$ vanishes when the Hamming
distance\break $d_{\rm H}(\sigma,\sigma')$ is larger than, say, $\delta.$
Starting from one given genotype\index{genotype}
$\sigma_0,$ after a certain number $\tau$
of evolutionary steps the genotype $\sigma(t)$ will be more
than $\delta$ away from $\sigma_0$: its fitness will be therefore
uncorrelated with $\sigma_0$ (except, of course, for the fact that
it is larger). Therefore, in the long run, the walk will
resemble a walk on a rugged fitness landscape,
apart from a rescaling of the elementary step length
from $\delta$ to one, and of the unit of time from $\tau$
generations to one.

One of the lessons to be taken from this result is that
the adaptative walk framework is too narrow to allow for
a high degree of adaptation, since the expected value
of fitness of local optima is so low. In order to explain
a higher degree of adaptation, one is led to introduce
mechanisms that allow to explore
larger regions of genotype\index{genotype space}
space. One possibility is the appearance of a ``hopeful
monster'',\index{hopeful monster} i.e., a  mutant
whose genotype is further
away from the dominant genotype than one or two
mutations. Another (already suggested by S.~Wright)
is the appearance of a chain of slightly unfit
mutants, one after the other, which may ``reach out''
for further fitness peaks. I do not find these
suggestions very convincing. However, these possibilities
cannot be discussed without a closer look at the genetic
structure of evolving populations.

\section{The quasispecies approach}\index{quasispecies}
There is no analytically treatable model, to my knowledge,
that describes fully the structure of a population evolving
on a nontrivial fitness landscape. In the Adaptative Walk
model, all genetic variability within the population
is neglected. The quasispecies model, introduced by
M.~Eigen\index{Eigen, M.} in the context of the theory of prebiotic
evolution \cite{Eigen71,Eigen88},
neglects fluctuations in the composition of
the population. In the case of nonoverlapping
generations that we consider here for simplicity,
it may be simply derived by introducing the effects
of mutation into eq.~(\ref{fund:eq}).\index{mutation}
Let us denote by $W(\sigma\leftarrow\sigma')$ the conditional
probability that, while attempting to reproduce
an individual of genotype\index{genotype} $\sigma',$ one produces instead
an individual of genotype $\sigma.$ Taking into account
this effect, the equation (\ref{fund:eq}) for the
number $n_{t+1}(\sigma)$ of individuals of genotype $\sigma$
at generation $t+1$ becomes the
quasispecies\index{quasispecies} (QS) equation:
\begin{equation}
n_{t+1}(\sigma)=\frac{1}{{\cal Z}_t}\sum_{\sigma'}
W(\sigma\leftarrow\sigma')A(\sigma')n_t(\sigma').
\label{QS:eq}
\end{equation}
The normalizing constant ${\cal Z}_t$ must be chosen in a
way to satisfy the external constraint imposed on the
population. The simplest constraint is constant population
size: $\sum_{\sigma}n_t(\sigma)=M={\rm const.}$, which implies
\begin{equation}
{\cal Z}_t=\frac{1}{M}\sum_{\sigma}A(\sigma)n_t(\sigma),
\end{equation}
where we have exploited the fact that $\sum_{\sigma}
W(\sigma\leftarrow\sigma')=1,$ $\forall\sigma'.$

This equation exposes its origin in the theory of chemical
reactions in that it neglects fluctuations in the
numbers $n_t(\sigma).$ This neglect is warranted
when these numbers are much larger than one, which is the case
when the different chemical species are few in number,
and the number of interacting molecules is very large. However,
in evolutionary theory, and even in the RNA replication
\index{RNA replication}
experiments discussed by W.~Gr\"uner in this meeting \cite{Gruner95},
\index{Gr\"uner, W.}
the number of points in genotype space is much larger
than the population size $M.$ It is possible nevertheless
to take it as a starting point, and we shall see that
it is valid at least in a particular
regime.

It is easy to derive the explicit form of the
mutation matrix $W(\sigma\leftarrow\sigma')$ when
one considers only point mutations with uniform
probability $\mu$:\index{mutation}
\begin{equation}
W(\sigma\leftarrow\sigma')=\mu^{d_{\rm H}(\sigma,\sigma')}
(1-\mu)^{N-d_{\rm H}(\sigma,\sigma')}.
\end{equation}

The Fundamental Theorem\index{Fundamental
Theorem of Natural Selection} is recovered in the limit $\mu\to 0.$
As soon as $\mu>0,$ however, the asymptotic composition of
the population is dictated by a balance of the effects of\index{selection}
selection and mutation not unlike the energy-entropy balance
determining the equilibrium in thermodynamics. It is
indeed possible to formulate the solution of the QS equation
in terms of equilibrium statistical mechanics\index{statistical mechanics}
\cite{Leuth87,Tara92}.\index{Leuth\"ausser, I.}
\index{Tarazona, P.}
The most interesting consequence of this analogy is the existence
of a phase transition between an ``ordered'' (selection-dominated)
regime and a ``disordered'' (mutation-dominated) one.
\index{phase transition} This transition has
been named the ``error threshold''.\index{error threshold}

Let us consider the ``single-peak landscape''\index{single-peak
landscape} defined by
\begin{equation}
A(\sigma)=\cases{A_0, &if $\sigma=\sigma^*;$\cr
A_1<A_0, &if $\sigma\neq\sigma^*.$}
\end{equation}
The maximum fitness\index{fitness} is reached for the isolated
sequence $\sigma^*,$ called the optimal or the master
sequence.\index{master sequence}
It is easy to solve numerically the QS equation by lumping
together all sequences as a function of their
Hamming\index{Hamming distance}
distance. For $\mu=0,$ we have $n_t(\sigma)\to n_{\infty}(\sigma)$
as $t\to\infty,$ where $n_{\infty}(\sigma^*)=M$ and
$n_{\infty}(\sigma)=0$ for $\sigma\neq\sigma^*.$
In other words, all genotype\index{genotype} sequences
in the population are identical, and equal to the master sequence..
When $\mu>0,$ the stationary distribution is sharply
peaked on the master sequence,\index{master sequence} but sequences within a
small
Hamming distance from it (close mutants) also appear with
nonnegligible frequency. This distribution of a master
sequence with its close mutants is called a {\em quasispecies\/}
\cite{Eigen71,Eigen88}.
\index{quasispecies}\index{Eigen, M.}

In this regime, the QS equation describes
rather faithfully the structure of the population.
Most of the genotypes are equal to the master sequence
or are close to it (in terms of the Hamming distance):
the frequency of these genotypes\index{genotype} is large enough for
the corresponding fluctuations to be negligible.
To be sure, further mutants appear and disappear
from the population: their frequency is small and
the relative fluctuations in $n_{\sigma}(t)$ are large.
However, they play essentially no role in the
dynamics.

As the mutation rate increases, the concentration of the master sequence
decreases. We can locate
the critical value $\mu^*$ of the mutation rate where the
master \index{master sequence}sequence concentration (as estimated from
first-order
perturbation theory) vanishes. It satisfies
\begin{equation}
\left(1-\mu^*\right)^N=\frac{A_1}{A_0}.\label{errthr}
\end{equation}
For $\mu>\mu^*,$ the population is no more ``hooked''
at the master sequence, and the QS equation predicts
an almost uniform concentration of all sequences.
Therefore $\mu^*$ is a good estimate of the location of
the error threshold. The error\index{error threshold}
threshold becomes sharper
and sharper as $N\to\infty,$ provided that the ratio
$A_0/A_1$ increases exponentially with $N.$

Beyond the error threshold, the predictions of the QS
equation cannot be taken at face value. In the usual
case in which the population size $M$ is much smaller
than the number of points in sequence space, $2^N,$
it is impossible to reach a stationary sequence distribution
with almost uniform concentration. One has instead a
wandering cloud of sequences, whose structure is dictated
by the reproduction-mutation mechanism, and where the effects
of selection can be neglected to a first approximation.
This regime is well described by the ``neutral theory''
\index{neutral theory}
due to M.~Kimura \cite{Kimura83}.\index{Kimura, M.}
It deserves a more thorough discussion, that is deferred to
the next section.

It is instructive to solve the QS equation in the rugged
fitness landscape\index{rugged fitness landscape} discussed
above \cite{Franz93}.\index{Franz, S.}\index{Sellitto, M.}
\index{Peliti, L.} One can
identify the error threshold
with a spin-glass\index{spin glass}
transition if one assumes that the ``typical''
values of the fitness behave like $\exp(N)$ for large $N.$
The role of the inverse temperature is played by\index{temperature}
$\beta=\frac{1}{2}\log(\mu/(1-\mu)).$ For $\beta>\beta^*,$
the population is essentially concentrated on a fitness\index{fitness}
optimum, while for $\beta<\beta^*$ all consequences of
selection disappear in the ``thermodynamic limit''.
It is therefore likely that the error threshold is
a general feature of all generic
fitness\index{fitness landscape} landscapes,
independently of their ruggedness.

The concept of the error threshold is central to the
theory of prebiotic evolution\index{prebiotic evolution}
\cite{Eigen71,Eigen88}.\index{Eigen, M.} A suggested mechanism for
the emergence of life is the formation of complex molecules
capable of self-reproduction. Given the accuracy $1-\mu$ of
the replication mechanism, eq.~(\ref{errthr}) sets an upper
bound to the length $N$ of these molecules. Reasonable
estimates of $\mu$ lead to values of $N$ which appear too short
to be able to start up a Darwinian evolutionary mechanism
eventually leading to the first cell. One way out of this problem
is to assume that the necessary biological information
was separated in several different molecules, each with
an $N$ smaller than the critical one, and related to one
another in a structure of chemical reactions like
the hypercycle \cite{Eigen79}\index{hypercycle} or
more general ones \cite{Szath87,Stadler93}.
\index{Szathm\'ary, E.} This problem may find a completely
different solution within the theory of
neutral networks\index{neutral networks}
expounded at this meeting by W.~Gr\"uner \cite{Gruner95,Schuster95}.

\section{Evolution in a flat
fitness\index{flat fitness landscape} landscape}
The relative weights of mutation and selection in
shaping the evolution of natural population has been
the subject of a hot debate since the
late sixties, when Crow and Kimura \index{Crow, J.}\index{Kimura, M.}
introduced the Neutral Theory of Molecular\index{neutral theory}
Evolution \cite{Crow70,Kimura83}. This theory
was prompted by the observation that natural
populations exhibit a much higher degree of genetic\index{variability}
variability at the molecular level than was previously
suspected. If selection were dominant, this would
imply that most of the variants which are found in
natural populations have {\em not yet\/} been
eliminated by Natural Selection. This, however, would mean
that actual populations have a much lower fitness\index{fitness}
than the optimal one.

Crow and Kimura suggested instead that most molecular
variants in the genotype have
the same fitness as the most common one. They
are therefore {\em selectively neutral}.\index{neutral theory} To be sure,
there {\em are\/} mutants corresponding to a much
smaller fitness that the dominant one, but they are
fast eliminated by Natural Selection, in accordance\index{Fundamental
Theorem of Natural Selection}
with the Fundamental Theorem. But the variability that
is left in the genotypes does not correspond to
a measurable effect on the fitness, again in accordance
with the Fundamental Theorem, which states that the
{\em fitness\/}\index{fitness} of all individuals (not their
\index{genotype}genotype)
is the same at stationarity. Evolution by increasing
\index{adaptation}adaptation,
in this view, is a comparatively rare phenomenon,
which has little bearing on the genetic structure
of the populations at the molecular level.

It becomes therefore rather interesting to describe
the structure of a population evolving in a flat fitness
\index{flat fitness landscape}landscape,
in which all genotypes have the same
fitness.

The results contained in this section are a translation
of the results of the Neutral Theory in the ``spin''
language which we have used so far \cite{Derrida91}.\index{Derrida, B.}
\index{Peliti, L.}
We consider a population of $M$ individuals, whose\index{genotype}
genotype $\sigma^{\alpha},$ $\alpha=1,2,\ldots,M,$
is identified by $N$ binary variables (units)
$\sigma_i^{\alpha}=\pm 1,$ $i=1,2,\ldots,N.$
Given the genetic structure $(\sigma^1,\sigma^2,\ldots,\sigma^M)$
at generation $t,$ the corresponding structure at
generation $t+1$ is obtained according to the following
procedure:
\begin{itemize}
\item[(i)] For each individual $\alpha$ of the new population,
one chooses, independently and with uniform probability,
the label $\alpha'=G_t(\alpha)$ of its parent among the
$M$ possible ones;
\item[(ii)] The genotype $\sigma^{\alpha}(t+1)$ is given by
$\left(\sigma^{\alpha}_1(t+1),\ldots,\sigma_N^{\alpha}(t+1)\right),$ where
\begin{equation}
\sigma_i^{\alpha}(t+1)=\epsilon_i^{\alpha}(t)
\sigma_i^{\alpha'}(t).\label{process}
\end{equation}
In this equation, $\epsilon_i^{\alpha}(t)=\pm 1$ is for each $i,$ $\alpha,$
and $t,$ an independent random variable with average
$\overline{\epsilon_i^{\alpha}(t)}={\rm e}^{-2\mu}.$
This equation defines the {\em bare mutation rate\/} $\mu.$
\end{itemize}
The reproduction process, by which each individual ``chooses''
its parent, is a random dynamical process applying
a $M$-point set into itself. This process has been thoroughly
studied by Derrida and Bessis
\cite{Derrida88},\index{Derrida, B.} and their
results have only to be translated into our language, to
obtain results on the statistics
of genealogies.

Let us fix our attention, for example, on the population at a given
generation much later than the beginning of the process.
Let us pick up at random $n$ individuals: it is a simple matter
to show that the probability $\pi_n$ that all $n$ individuals
have $n$ different parents is given by
\begin{equation}
\pi_n=\left(1-\frac{1}{M}\right)\left(1-\frac{2}{M}\right)
\ldots\left(1-\frac{n-1}{M}\right)\simeq1-\frac{n(n-1)}{2M},
\end{equation}
assuming $n\ll M.$
The probability $p_n(t)$ that each of the $n$ individuals had a different
\index{ancestor}ancestor $t$ generations ago is obviously given by
\begin{equation}
p_n(t)=\pi_n^t\simeq\exp\left(-\frac{n(n-1)t}{2M}\right).
\end{equation}

Let us now say that two individuals belong to the same
$t$-family if they had the same ancestor $t$ generations ago.
The number $F(t)$ of $t$-families is a random variable,
which changes as the population evolves. Let us
fix again our attention on the population at a given
time, and consider $F(t)$ as a function of $t.$
This number is reduced, as $t\to t+1,$
if the ancestors of two different $t$-families
had the same parent. This happens with a probability
equal to $1-\pi_{F(t)}=F(t)(F(t)-1)/(2M).$ We can
thus write down a \index{mean field}``mean field equation'' for the
{\em average\/} $\Phi(t)=\overline{F(t)}$:
\begin{equation}
\frac{d\Phi(t)}{dt}=-\frac{\Phi(t)(\Phi(t)-1)}{2M}.
\end{equation}
The solution of this equation reads
\begin{equation}
\Phi(t)=\left(1-{\rm e}^{-t/2M}\right)^{-1}.
\end{equation}
Therefore, after a number of generations
essentially equal to $M,$ all individuals
in the population share the same \index{ancestor}ancestor.
Derrida and Bessis \cite{Derrida88} have calculated
the probability $Z_k(t)$ that $F(t)=k.$ One can then obtain
an exact expression for $\Phi(t)$:
\begin{equation}
\Phi(t)=
\sum_{k=0}^{\infty}k\,Z_k(t)=
\sum_{k=1}^{\infty}(2k-1)\exp\left[-\frac{k(k-1)t}{2M}
\right].
\end{equation}
This expression agrees with the mean-field one when $t\ll M,$
yielding $\Phi(t)\simeq 2M/t,$ but deviates from it in the
fluctuation-dominated range $t\ge M,$ where $F(t)\simeq 1.$
It is actually possible to compute the probability distribution
of the sizes of all $t$-families, obtaining the
result that all possible ways of breaking the population
of $M$ individuals into $k$ $t$-families have the same
probability, once $k$ is given. As a result, the sizes of
$t$-families fluctuate wildly.

It is indeed possible to calculate more explicitly
the distribution of the
\index{genetic structure}genetic structure of the population.
Let us remark first of all than in the infinite
genotype limit $N\to\infty,$ the genetic overlap
$q^{\alpha\beta}$ between two individuals $\alpha$
and $\beta$ is a function of their relatedness.
If the last common \index{ancestor}ancestor of $\alpha$
and $\beta$ had existed $\tau^{\alpha\beta}$
generations before the present,
one would have $q^{\alpha\beta}=\exp(-4\mu\tau^{\alpha\beta}).$
In fact, the \index{genotype}genotypes of the two independent lineages
of ancestors of the two individuals have performed two independent
random walks on the hypercube,
with an average rate of $\mu N$
steps per generation.

Therefore the distribution function $P(q)=\av{\delta(q^{\alpha\beta}
-q)}$ of the \index{overlap}overlap reflects the genealogical structure
of the population~\cite{DH91}.\index{Derrida, B.}\index{Higgs, P. G.}
 At any given time, a peak in $P(q)$
represents a subpopulation of $\nu$ individuals whose last
common \index{ancestor}ancestor existed $\tau$ generations ago:
the location of the peak is given by $\exp(-4\mu\tau),$
while its height is proportional to $\nu^2.$
As time goes on, these peaks move towards zero,
according to the law just stated,
while their height fluctuates. From time to time
some of the peaks disappear, and new ones arise from the
$q\sim 1$ region, as new subpopulations appear.
The genetic structure of the population is therefore
a stochastic process, which evolves in time according to the
particular realization of the mapping $\alpha\to G_t(\alpha)$
of each individual to its parent. One must therefore
distinguish two kinds of averages \cite{Derrida91}:
\begin{itemize}
\item The \index{population average}{\em population average\/}
denoted by $\av{\ldots}$: for
example, the average overlap $Q=\av{q}$ in the population is defined by
\begin{equation}
Q=\av{q}=\left[\left(\begin{array}{c}M\\2\end{array}
\right)\right]^{-1}\sum_{(\alpha,\beta)}
q^{\alpha\beta},
\end{equation}
where the sum runs over all different pairs $(\alpha,\beta)$
of individuals in the population.
\item The \index{process average}{\em process average\/}
denoted by $\overline{\strut\ldots}.$
One has for example \cite{Stewart76,Derrida91}:
\begin{eqnarray}
\overline{Q}&=&\overline{\av{q}}=\frac{1}{1+\lambda};\\
\overline{Q^2}&=&\overline{\av{q}^2}=
\frac{\lambda^2(9\lambda^2+18\lambda+4)}{(\lambda+1)
(\lambda+2)(3\lambda+1)(3\lambda+2)},
\end{eqnarray}
where we have introduced the notation $\lambda=1/(4\mu M).$
\end{itemize}
It is obvious from the fact that $\overline{Q^2}>\overline{Q}^2$
that $Q$ is itself a random quantity. This observation implies that
the genetic structure of any given sample, even of a very
large population, will be in general very different from the
average one: and therefore that predictions based on a
\index{mean field}``mean
field'' approach, like the QS equations, could be rather misleading.

The average \index{overlap}overlap $\overline{Q}$ depends on population
size $M,$ and decreases as $M$ increases. Thus
genetic \index{variability}variability
increases with increasing population size.
In most natural populations genetic \index{variability}variability is
much smaller than expected on the basis of this result
\cite{Kimura83}.\index{Kimura, M.} In fact, natural populations are
often the outcome of a comparatively recent ``population boom''
involving a rather small \index{founder population}founder
population. In order to
reach the ``equilibrium'' value quoted above one should
wait of the order of $M$ generations, where $M$ is population
size. This is often too long, and the result is that the
actual \index{variability}variability
reflects the much smaller size of the founder population.

It is also interesting to monitor the evolution of the
{\em average \index{genotype}genotype\/}
\begin{equation}
\av{\sigma}=\left(\av{\sigma_1},\ldots,\av{\sigma_N}
\right).
\end{equation}
The genetic \index{drift}drift of the population (a physicist would
call it diffusion) is represented by the correlation function
\begin{equation}
K(t)=\frac{1}{N}\sum_{i=1}^N
\overline{\av{\sigma_i}_{t_0}\av{\sigma_i}}_{t_0+t},
\end{equation}
where $\av{\ldots}_t$ denotes the
population average at generation $t.$
The exponential decrease of the
\index{correlation}correlation $K(t)\propto
\exp(-2\mu^*|t|)$ defines the {\em effective mutation
rate\/} $\mu^*.$ A simple calculation from eq.~(\ref{process})
yields
\begin{equation}
K(t)=Q\exp(-4\mu|t|).
\end{equation}
In our case therefore, the
\index{effective mutation rate}effective mutation rate is equal
to the bare one, and in particular, it is independent of population
size. This rather surprising result is known as the \index{Kimura, M.}
Kimura theorem~\cite{Kimura83}.

The previous result hold almost without
change if all fit \index{genotype}genotypes have the
same \index{fitness}fitness value,
while unfit ones have a negligible one.
Let us assume that a fraction $x$ of the genotypes is unfit,
and therefore practically unable to reproduce, and that fit
and unfit genotypes are distributed at random on the hypercube.
Therefore $Nx$ neighbors of every fit
genotype will be unfit on average. If a
\index{mutation}mutation appears, it
will lead to an unfit genotype with probability $x.$
It will be safer not to mutate, since one's parent is
fit by definition. Therefore the effective mutation rate
$\mu^*$ will be smaller than the bare one $\mu,$ and given
approximately by $\mu^*\simeq\mu(1-x)$~\cite{Kimura83,Bastolla92}.

The effective mutation rate $\mu^*$ can only be nonzero if
the clusters of fit genotypes span the hypercube, i.e., if
it is possible to connect two arbitrarily different
fit genotypes via a chain of fit mutants. Let us say that
the fit genotype\index{genotype}
$\sigma$ belongs to the same cluster as the fit genotype
$\sigma'$ if it differs in only one unit $\sigma_i$
either from $\sigma'$ or from a fit genotype which belongs
to the same cluster of $\sigma'.$ If $x$ is small,
there is a large cluster of fit genotypes which spans
the hypercube: starting from any point on it, one can reach genotypes
whose overlap with the initial point is arbitrarily small.
However, when $x>x^*\sim 1-1/N,$ the space of fit genotypes breaks
down into small clusters, and it is never possible to wander far
away from the initial point stepping only on fit genotypes
via single-point mutations. In this case memory of the initial genotype
is maintained forever. This phenomenon is called the percolation
\index{percolation}threshold, and we see that it is closely related to the
\index{error threshold}error threshold discussed in the previous section.
We can introduce the order parameter
\begin{equation}
q_{\infty}=\lim_{t\to\infty}\frac{1}{N}\sum_i
\overline{\av{\sigma_i(t)}}^2,
\end{equation}
where it is understood to take the process average with
a fixed initial genotype $\sigma(0).$ This
\index{order parameter}order parameter is nonzero
in the ``trapped'' regime $x>x^*.$

In the ``wandering'' regime $x<x^*,$ the population evolves
in a {\em neutral network},
\index{neutral network} and is able to explore larger and larger
regions of sequence space as time goes
on~\cite{Gruner95,Schuster95}.\index{Gr\"uner, W.}\index{Schuster, P.}
At any given point, the
number of fit mutants can be small, but the sequence space has
a large connectivity, and the neutral network can efficiently
span it.

In a number of \index{protein}proteins
one can relate the number of aminoacid\index{aminoacid}
substitutions in different {\em taxa\/} to their
respective time of divergence, i.e., the time of the existence of their
last common \index{ancestor}ancestor. One obtains a
well-defined substitution
rate which is specific of the protein, except for very
conservative proteins like histones~\cite{Ayala76}. This suggests
that proteins also evolve along
\index{neutral network}neutral networks:
unfit aminoacid substitutions are eliminated at each step, but
fit substitutions, which are selectively neutral, are
retained in accordance with the
\index{neutral theory}neutral theory~\cite{Kimura83}.

\section{Two-parent reproduction and the
\index{species}origin of species}
Since most of the organisms which are close to
us in daily life reproduce \index{sex}sexually, it is interesting
to ask if the results of the previous section
hold for a two-parent reproduction \index{Serva, M.}\index{Derrida, B.}
\index{Higgs, P. G.}
mechanism~\cite{Serva91,DH91}.
The model can be defined in analogy with eq.~(\ref{process}).
One chooses, at each generation and for each individual
$\alpha,$ {\em two\/} parents $\alpha'$ and $\alpha^{\prime\prime}.$
The \index{genotype}genotype $\sigma^{\alpha}(t+1)$ of the individual
$\alpha$ in the new generation is then given (for $i=1,2,\ldots,N$) by
\begin{equation}
\sigma^{\alpha}_i(t+1)=\epsilon_i^{\alpha}(t)
\left[\xi_i^{\alpha}(t)\sigma_i^{\alpha'}(t)+
\left(1-\xi_i^{\alpha}(t)\right)
\sigma_i^{\alpha^{\prime\prime}}(t)\right].
\end{equation}
Here $\epsilon_i^{\alpha}(t)=\pm 1$ represents the effects of
\index{mutation}mutations
as in the previous section, while $\xi_i^{\alpha}(t)
\in\{0,1\}$ determines from which parent our individual inherits
the state of its unit $\sigma_i$: if $\xi_i^{\alpha}=0,$ from
$\alpha';$ otherwise, from $\alpha^{\prime\prime}.$ One
has $\overline{\xi_i^{\alpha}(t)}=\frac{1}{2}.$

The genetic variability\index{variability} is again
expressed in terms of the average genetic overlap\index{overlap}
\begin{equation}
A=\overline{Q}=\overline{\av{q}}=\overline{\sigma_i^{\alpha}
\sigma_i{\beta}}.
\end{equation}
One obtains for $\overline{Q}$ the equation
\begin{equation}
\overline{Q}={\rm e}^{-4\mu}\frac{1}{4}
\left[\frac{4}{M}+4\left(1-\frac{1}{M}\right)
\overline{Q}\right].
\end{equation}
This equation implies $\overline{Q}=\lambda/(1+\lambda),$ as
for the one-parent reproduction mechanism. Fluctuations
are determined by the quantities
\begin{eqnarray}
B&=&\overline{\av{q^2}}=\overline{\sigma_i^{\alpha}\sigma_i^{\beta}
\sigma_j^{\alpha}\sigma_j^{\beta}};\\
C&=&\overline{Q^2}=\overline{\av{q}^2}=\overline{\sigma_i^{\alpha}
\sigma_i^{\beta}\sigma_j^{\gamma}\sigma_j^{\delta}};\\
D&=&\overline{\sigma_i^{\alpha}\sigma_i^{\beta}\sigma_j^{\alpha}
\sigma_j^{\gamma}}.
\end{eqnarray}
It is understood that different indices take on different values.
We assume $M\to\infty,$ but $4\mu M\to\lambda^{-1}={\rm const.}$
In this limit, we have
\begin{eqnarray}
3B&=&C+2D;\\
\left(\frac{1}{\lambda}+3\right)C&=&A+2D;\\
4D&=&2D+2C+A.
\end{eqnarray}
These equations imply
\begin{equation}
A^2=B=C=D=\frac{1}{(1+\lambda)^2},
\end{equation}
and therefore
\begin{equation}
\overline{\av{q}^2}=\overline{\av{q^2}}=\overline{\av{q}}^2.
\end{equation}

Stated in other words, this result implies that in
a population in which all possible pairs have the same
probability of producing offspring (such populations are
called {\em panmictic\/}), the genetic difference
between any two individuals has the same value with probability
one. To be sure, this result has to be modified in
actual populations: siblings {\em are\/} genetically more
similar than strangers; however it is in agreement with
the fact that specific genetic correlations are lost
{\em very\/} rapidly as the genealogical relatedness
decreases.

Higgs and Derrida~\cite{DH91} have taken advantage of this
result to introduce a minimal model for the formation of
biological species. They assume that a pair of individuals can
produce offspring only if their genotypes are not too
different: more explicitly, if their genetic overlap
$q$ is larger than a threshold $q_0$ (fecundity threshold).
As long as $M$ and $\mu$ are such that $\overline{Q}$
(as obtained above) is larger than $q_0,$ nothing happens;
but if the population $M,$ or the mutation rate $\mu$ is too large,
the population splits into several subpopulations:
the genetic overlap $q$ of
individuals belonging to the same subpopulation is larger than
$q_0,$ whereas the overlap belonging to different
subpopulations is smaller. Therefore, offspring can only be
produced by pairs of individuals belonging to the
{\em same\/} subpopulation. This is exactly the definition
of the biological species, according to Mayr~\cite{Mayr42}:
\begin{quote}
Species are groups of actually or potentially interbreeding
natural populations which are reproductively isolated from
other such groups.
\end{quote}
The actual behavior of the population is extremely irregular:
the size of the subpopulations fluctuates according to the
irregularities in their sampling, much in the same way as family
size in the one-parent reproduction mechanism; and the
value of the corresponding characteristic overlap fluctuates
in agreement with the expression obtained above which relates
the genetic overlap of a panmictic population to its size.
{}From time to time, a subpopulation becomes too large, and the
corresponding overlap hits the threshold. After a short
period of ``confusion'' two new subpopulations (species)
arise. The mutual overlap between different species evolves
in time according to the exponential law that we have
derived in the previous section.
The process average of the overlap $Q$ over the whole population
is, rather remarkably, given by the same expression valid
for a panmictic population \cite{DH91}. There has not yet
been an explanation of this result which appears
very clearly in the populations.

The same model can be generalized to treat the effects
of geographic isolation~\cite{Manzo94}.
One considers a population with reproduces with the
same two-parent mechanism discussed above, but which is
distributed in several ``islands''.
Reproducing pairs can only be formed among individuals
inhabiting the same island, but before each ``mating season''
an individual can move from one island to a neighboring
one with a small probability $\epsilon.$
As a consequence, the overlap between individuals
belonging to the same island is larger than between
different islands. When the migration rate becomes so
small that the average overlap between neighboring
islands drops below the fecundity threshold, the system
start behaving in the same irregular way as discussed
above. Again, the {\em average\/} overlaps (either between
islands or within one island) behave as in the absence of
the fecundity threshold.

There is an interesting phenomenon which takes place in
rings of islands. One may reach a regime in which the
average overlap between neighboring islands is above threshold,
while that between islands further away is below.
In this case it is possible to start from one islands and to
move in one direction in the ring, always finding
mutually fecund populations in neighboring islands.
However, coming back to the starting point, one finds
a reproductive barrier, and possibly the coexistence,
in the same islands, of two populations mutually
sterile. This phenomenon appears rather frequently
in the simulations \cite{Manzo94} and can be related to
the {\em circular invasion\/} phenomenon observed
in natural populations~\cite{Mayr63}. For example, the northern
skylark {\sl Larus argentatus\/} exhibits a group
of populations which are mutually interfecund if one
starts from Scandinavia and moves towards the East,
reaching, over Siberia, to North America. However,
where the last American population overlaps with
the first European one, they are no more mutually
interfecund. Therefore the relation ``being mutually
interfecund'' is not necessarily an equivalence relation.

If one goes over to consider all-or-none selection
in the presence of a two-parent
reproduction mechanism\index{Peliti, L.}
\cite{Peliti94},\index{Bastolla, U.} one observes another peculiar
phenomenon. The effective mutation rate $\mu^*$
does not appear to depend on the fraction $x$ of unfit
genotypes, as long as $x$ is small. When $x$ crosses
a threshold $x^*$ (which increases with the genome
size $N$) the mutation rate drops suddenly and the
average fitness of the population increases. This
behavior can be understood in terms of collective
adaptation, i.e., of the search for a region which
optimizes the {\em recombinational fitness\/} of the individual,
a quantity which measures also the likelihood
that the offsprings of an individual, obtained by
mating with the other members of the population, are
fit. A quantitative theory of this transition (which
is rather striking in the simulations) is still lacking.

\section{Two-level selection and the maintenance of unselfish
genes}
The last observation prompts us to consider situations
in which the selection process leads to potential
conflicts. One case, much studied in the literature,
is the possible existence of {\em
\index{unselfish gene}unselfish genes},\index{altruism}
which determine, in the individual which expresses them,
a behavior disadvantageous to the carrier, but
beneficial to the group to which it belongs.
While the existence of such genes has not yet been
proved beyond doubt, it is an interesting problem
in itself to see whether the interaction between the
two selection levels, the individual and the group,
can lead to the permanence of these genes in the population.
Here I shall only report briefly a model which has
been introduced and analyzed by R.~Donato,\index{Donato, R.}
M.~Serva\index{Serva, M.} and
myself\index{Peliti, L.}~\cite{Donato95,DSP95}.

We consider a population divided into groups
\index{deme}(demes) of $L$
individuals. Mating is only allowed within a group.
Heredity acts according to the usual Mendelian\index{Mendel, G.}
mechanism.
There is a behavioral locus with two alleles: a selfish
(S) allele (dominant), and an unselfish (U) one (recessive).
(It is easy to modify the model to consider more alleles,
different reproduction schemes, etc.) The fitness of an individual
depends on two factors: (i) if it is unselfish homozygote
(UU) it is {\em reduced}, by a factor $(1-r),$ with respect
to the other members of a group; (ii) if it belongs to a group
with a large enough fraction $x$ of
UU-individuals (larger than a threshold $x^*$), it is {\em enhanced},
by a factor $(1+c),$ with respect to groups
which do not satisfy this condition.

This definition can be summarized in the following table:
\begin{table}[h]
\begin{center}
\caption{Fitness table}
\begin{tabular}{lll}
\hline
&$x<x^*$&$x \ge x^*$\\
\hline
Genotype: UU&$1-r$&$(1-r)(1+c)$\\
Other genotypes&1&$1+c$\\
\hline
\end{tabular}
\end{center}
\end{table}

As a consequence of this selection scheme, the fraction
$x$ of unselfish individuals decreases within any given
deme. On the other hand, demes with $x>x^*$ have higher
average fitness and tend to expand at the expenses of
the others. When a \index{deme}deme grows too large, it splits
into two demes, and its members are redistributed at
random between the two new demes. We can represent
this process via a ``deme \index{fitness}fitness'' $A(x),$
proportional to the total
fitness of individuals which form each deme, and
given by
\begin{equation}
A(x)=\cases{1-x+x(1-r),&if $x<x^*;$\cr
(1+c)\left[1-x+x(1-r)\right],&if $x\ge x^*.$}
\end{equation}

In the limit of very large population (infinite number of demes)
the process can be described by a quasispecies\index{quasispecies}
equation at the level of demes. Denoting by $\rho_t(x)$
the fraction of demes with a fraction
$x$ of unselfish\index{unselfish gene}
individuals, we have
\begin{equation}
\rho_{t+1}(x)=\frac{1}{{\cal Z}_t}\int_0^1 dx' P(x\leftarrow x')
A(x')\rho_t(x').
\end{equation}
In this equation, $P(x\leftarrow x')$ is the conditional probability
density to produce a deme with a fraction $x$ of unselfish individuals,
starting from one with a fraction $x'.$ This probability contains two
effects: (i) the systematic decrease of $x,$ due to the
disadvantage of \index{altruism}altruism; (ii) the fluctuations due to random
sampling, due to the finite deme size $L.$

The quasispecies\index{quasispecies} equation for
the demes can be solved numerically
for the steady state distribution. One finds a line
$(1-r)(1+c)=f(L)$ which separates two regimes: when $r$
is too large (or $c$ is too small), the distribution is
peaked at $x=0$, and therefore unselfish genes are wiped out
of the population; otherwise, the distribution is nontrivial,
and the average value of $x$ is different from zero. The transition
appears to be discontinuous. It is interesting to remark that
it is the competition between demes that keeps unselfish genes
in the population: there is no ``optimal value'' for $x.$
Again, the ``steady state'' hides a complicated dynamical behavior:
demes with high values of $x$ increase in number, but their
values of $x$ decrease; however, new ones with high values
of $x$ arise from the splitting of old ones, and so on.

\section{Conclusions}
Fitness as an individual property, in the way we have used
here, is a powerful tool for model building. However, it
is not measurable in the field, because the actual number
of viable offspring of an individual is the outcome of its
complex interaction with other members of the same population
and with its environment. Some aspects of these interactions,
from the point of view of evolutionary success, can be captured
by the game\index{game theory} theory
approach~\cite{Sigmund93}.\index{Sigmund, K.} The most
important aspect is the difference between local optimization,
i.e., fitness optimization at the level of individual,
and global optimization, at the level of community.

Community can be formed by members of the same species, or
of different species. One of the key problems in understanding
evolutionary innovation is the evolution of
\index{individuality}individuality~\cite{Buss},\index{Buss, L.} i.e.,
of the organization of different units into a single integrated
organism to which it is possible to assign a fitness. This
is also the problem of the emergence of
\index{mutualism}mutualism~\cite{Duch95} and can be
the key point in the understanding of the
evolution of multicellular organisms.

Nevertheless the concept of fitness, with its strong aspects
of ``physicalism'' related to its similarity with energy,
is a very convenient stepping stone to enter, as physicists,
in the arena of evolutionary theory.

\section*{Acknowledgments}
I warmly thank my collaborators: U.~Bastolla, B.~Derrida, R.~Donato,\break
F.~Man\-zo, M.~Sellitto and especially M.~Serva. I also thank
W.~Fontana for much-needed encouragement and illumination.
The support of the INFN Sections of Napoli and Roma made
possible much of my work in this area.
And I am grateful to the organizers
of the NATO ASI on Physics of Biomaterials, and in particular to
Professor T.~Riste, for having given me the opportunity for
expressing my point of view on evolutionary theory.

\end{document}